\pdfoutput=1
\documentclass[twocolumn,showpacs,preprintnumbers,amsmath,amssymb, prl]{revtex4}

\usepackage{dcolumn}% Align table columns on decimal point
\usepackage{bm}% bold math
\usepackage{hyperref}
\usepackage{amsmath}
\usepackage{amsfonts}
\usepackage{graphics}
\usepackage{psfrag}
\usepackage{graphicx}
\usepackage{epsfig}
\usepackage{rotating}
 \usepackage[latin1]{inputenc}
 \usepackage{color}
 \usepackage{rotating}
 \usepackage{threeparttable}  % Revisions added by Zunfeng Wang for table adding Apr. 29, 2011
 \usepackage{multirow}        % Revisions added by Zunfeng Wang for table adding Apr. 30, 2011
 \usepackage{color}
\usepackage{lscape}

\definecolor{DarkGreen}{rgb}{0,0.5,0}
\definecolor{Grey}{rgb}{0.5,0.5,0.5}
\definecolor{DarkYellow}{rgb}{1,0.7,0}
\definecolor{Violet}{rgb}{0.6,0.0,0.7}
\definecolor{Brown}{rgb}{0.5,0.3,0}

\makeatletter

\newcommand{\Rmnum}[1]{\expandafter\@slowromancap\romannumeral #1@}
\makeatother

\begin{document}

%\preprint{LMU-ASC XX/08}

\title{Test the Principle of Maximum Entropy in Constant Sum 2 $\times$ 2 Game:\\Evidence in Experimental Economics }
\author{Bin Xu$^{1,2}$}
\author{Hongen Zhang$^{3}$}
\author{Zhijian Wang$^{1}$}
\author{Jianbo Zhang$^{3}$\footnote{E-mail: jbzhang08@zju.edu.cn}}

\affiliation{$^{1}$Experimental Social Science Laboratory, Zhejiang University, Hangzhou, 310058, China}
\affiliation{$^{2}$Public Administration College, Zhejiang Gongshang University, Hangzhou, 310018, China}
\affiliation{$^{3}$Department of physics, Zhejiang University, Hangzhou, 310027, China}

%http://www.mendeley.com/research/entropy-and-entropy-production-in-simple-stochastic-models/

%\email{tobias.reichenbach@physik.lmu.de}
%\affiliation{%

\date{\today}% It is always \today, today,
\begin{abstract}
Entropy serves as a central observable which indicates uncertainty  in many chemical, thermodynamical, biological and ecological systems, and the principle of the maximum entropy (MaxEnt) is widely supported in natural science. Recently, entropy is employed to describe the social system in which human subjects are interacted with each other, but the principle of the maximum entropy has never been reported from this field empirically. By using laboratory experimental data, we test the uncertainty of strategy type in various competing environments with two person constant sum $2 \times 2$ game. Empirical evidence shows that, in this competing game environment, the outcome of human's decision-making obeys the principle of maximum entropy.

\end{abstract}

\maketitle
%\textcolor[rgb]{0.14,0.51,0.11}{ggg}

\section{1 Introduction}
Entropy is a concept from physics. Shannon~\cite{Shannon1948}(1948) introduced entropy to information theory which is used to measure the uncertainty of a system. It serves as a central observable in many chemical, thermodynamical, biological and ecological systems. The maximum entropy principle~\cite{Jaynes1957} introduced by E.T.Jaynes$(1957)$ indicated that random crossing is an irreversible process, in which the entropy increases to its' maximum value.  This principle is widely supported in natural science. Recently, Bednary et.al.~\cite{Yan2011} and Cason et.al.~\cite{Cason2009} apply the standard entropy concept to laboratory social interaction system as the signal of cognitive complexity. But the principle of maximum entropy has never been reported from social science empirically.

In game theory, the social system can be abstracted as the process of individual decision-making and the outcome or social state. While a specific game is given, the potential social states are determined. This implicates that one can measure the entropy of the social system and test the principle of the maximum entropy in this system. On the other hand, experimental economics brought theoretical game to laboratory  to test the  prediction of game theory. Therefore, it allows us to test the principle of maximum entropy in social system by using economics experimental data.

Constant Sum Game is one of the classical games implying conflict behavior in a competing  environment which has Mixed Strategy Nash Equilibrium(MSNE). That means there should exist some uncertainty in decision-making in the game, and in consistent with this uncertainty, there should be some probability distribution of social states in repeated games. Obviously, the probability distribution should not be the same under different constraint conditions, i.e., different payoff matrices. Our question is, giving the constraint condition, does this competing game system obey the principle of maximum entropy? If it does, how to test it?

In this paper, using a series of economics experimental data of constant sum games, we firstly test the principle of the maximum entropy in human interaction social system. The paper is organized as follow: section two describes the entropy of an experimental game system in formal; section three provides the prediction of the principle of maximum entropy in constant sum game system under different payoff matrices; section four introduces the principle of experimental data selection; section five demonstrates the result; finally, conclusion and discussion.

\section{2 Entropy in laboratory social system}

\begin{figure}
\centering
\includegraphics[angle=0,width=3.5cm]{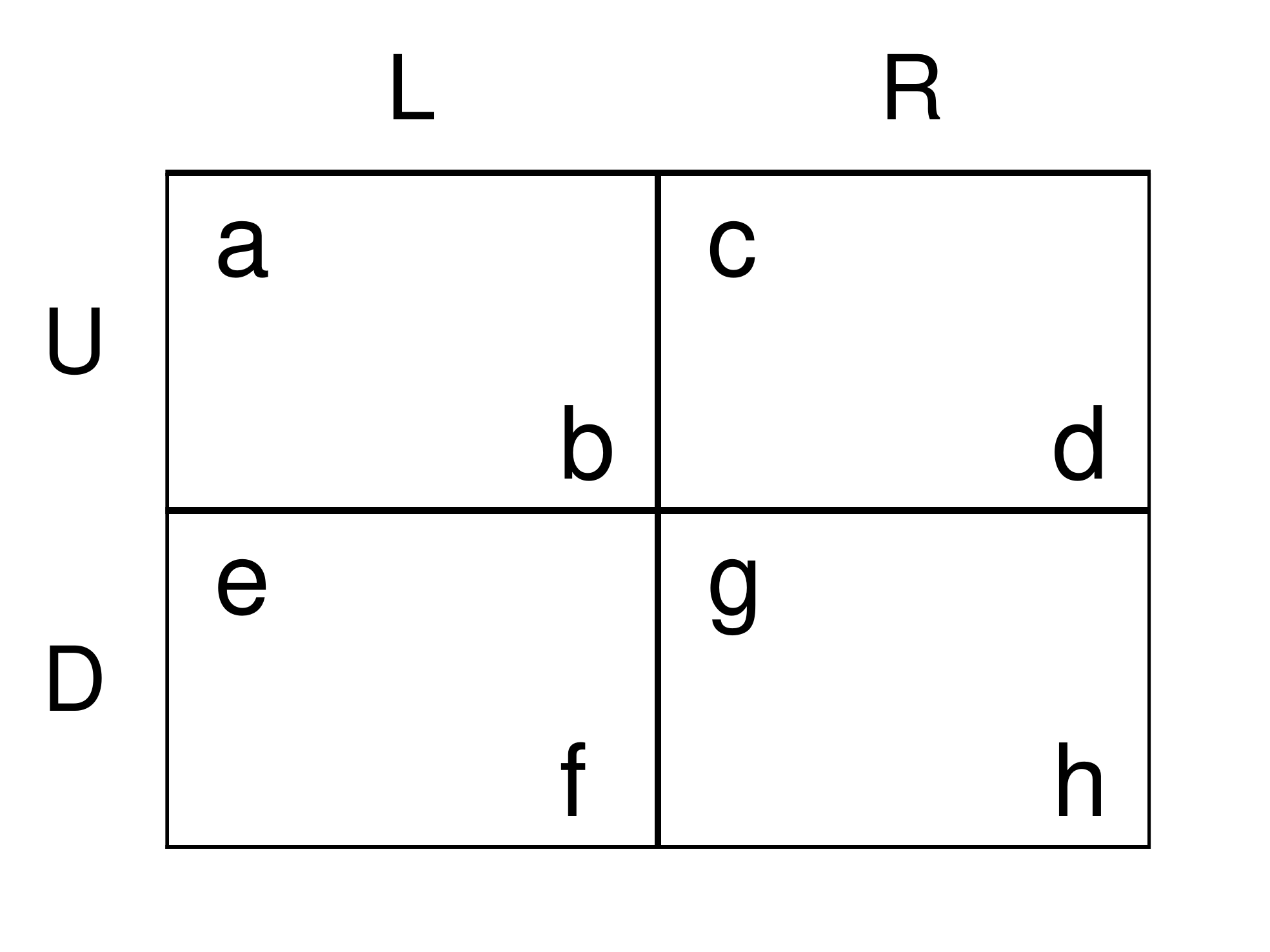}
\caption{Payoff Matrix.  Player 1's payoff in the upper-left corner, player 2's payoff in the down-right corner.
\label{fig:payoffmatrix}}
\end{figure}

For testing the entropy of the social system, we begin this task with its' simplest and most feasible type. Constant Sum 2 $\times$ 2 Game is one of the simplest games which indicates social interaction in competing environment, and allows us to conduct this game appropriate in experimental laboratory by using human subjects. 2 $\times$ 2 Game that is usually described in its' formal form as in Fig.~\ref{fig:payoffmatrix}.
%\begin{table}[htbp2]
%\small
%\caption{\label{tab:PayoffMatrix} Payoff Matrix of $2 \times 2$ game}
%\centering
%\begin{tabular}{cccccccc}
%
%  &           &&        &player&2 \\
%
%\hline
%  &           &&       & $L$     & $R$ \\
%  & player 1  &&   $U$ & $a,b$ & $c,d$ \\
%  &           &&   $D$ & $e,f$ & $g,h$ \\
%\hline
%\end{tabular}
%\end{table}

 The player $1$ and player $2$ are the two persons of the game who each have two strategies. The strategies for player $1$ are $U$ and $D$, where $U$ means player 1 chooses up and $D$ down, and for player $2$ are $L$ and $R$, where $L$ means left and $R$ means right. The numeral $a$ is the payoff for player $1$ when she/he chooses $U$ and the player $2$ chooses $L$, and the $b$ is for player $2$ in the same pair, and the rest is in the same manner. Constant Sum means the payoff sum in each cell is equal. This implicates that one would get less if the other gets more.

 Individual stage game strategies have been modeled~\cite{Yan2011} as a discrete random variable $X$. In the case of this paper, the variable $X$ may take on the values $x$ $\in \textit{X(i,j)}$ . Here \textit{X(i,j)} is the finite set of all possible values for $X$, ${i=u,d}$,  ${j=l,r}$, in other words, $x\in \textit{X(i,j)= \{(u,l),(u,r),(d,l),(d,r)\}}$ indicate all possible individual stage game strategies. The probability that $X$ takes on the particular value $x$ is written as $Pr(X = x)$, or just $p(x)$. The probability $p(x)$ is referred to the joint probability that player 1 chooses \emph{i} meanwhile player 2 chooses \emph{j} which is named the social state of strategies$(i,j)$ in these repeated games. The entropy of a random variable $X$ with a probability density function, ${p(x) = Pr(X = x)}$, is defined by
\begin{equation}
S(X)=-\sum_x p(x) \log_{2} p(x),
\label{eq:entropydefine}
\end{equation}
which is used to measure the amount of stochastic variation in a random variable that can assume a
finite set of values. Therefore, we come up with some mathematical entity that takes a probability
distribution and returns a number that can be interpreted as a measure of the uncertainty associated with that distribution. When using logarithms of base $r=2$, that measure is expressed in binary variables
(bits).

 Throughout the analysis, we use the convention that ${0 log 0 = 0}$. The entropy in a generic $2 \times 2$ game is in the interval $[0, 2]$, with the lower bound indicating certainty, i.e., all outcomes are more likely in one social state, and the upper bound indicating a uniform distribution among the four social states. A uniform distribution corresponds to complete uncertainty, because in that case, every state is equally likely to occur.
 %, you can't get more uncertain than that.

% The cause of behavioral variation could be strategic uncertainty over what the other player will do~\cite{feldman1997}.

\section{3 prediction of the principle of maximum entropy}

The uncertainty may not be equal under different payoff matrices. The maximum entropy itself would be different among games. In other words, distribution $p(u)$ and $p(l)$ determines the maximum entropy.

Given a certain distribution $p(u)$ and $p(l)$, the principle of maximum entropy indicates that the random variable $X$ should reach the maximum uncertainty. That is to say, the entropy is maximized by independent random variables.

There are four variables to describe the joint distribution, $p(u,l)$, $p(u,r)$, $p(d,l)$ and $p(d,r)$ in these games. Supposing the mean observations $p(u)$ and $p(l)$ have been obtained,  there are three constraints for the four variables,
%\begin{eqnarray}\label{eq:concentrate}
    $p(u,l)$ + $p(u,r)$ = $p(u)$,
    $p(u,l)$ + $p(d,l)$ = $p(l)$ and
    $p(u,l)$ + $p(u,r)$ + $p(d,l)$ + $p(d,r)$ = 1.
%\end{eqnarray}

%It will be possible to go through all the steps analytically.
We expressed all of the unknown probabilities in terms of the fourth $p(d,r)$. Therefore, these expressions were substituted into the formula for entropy $S(X)$, so that it was expressed in terms of a single probability. Thus, the entropy in Eq.~(\ref{eq:entropydefine}) depends on only one variable, $p(d,r)$.  Let
%\begin{eqnarray}
%  p(d,l) &=& 1-p(u)-p(d,r) \nonumber\\
%  p(u,l) &=& p(l)-1+p(u)+p(d,r) \nonumber\\
%  p(u,r) &=& 1-p(l) - p(d,r)
%  \label{eq:concentrateArray}
%\end{eqnarray}
% \begin{equation}
$dS/dp(d,r)$=0,
%\label{eq:entropy3}
%\end{equation}
% we get
%
% \begin{eqnarray}
%  & &\log_2\!\left(p(d,r)\right) \nonumber\\
% &+& \log_2\!\left(p(u) + p(l) + p(d,r)  - 1\right)\nonumber\\
% &-& \log_2\!\left(1 - p(d,r) - p(u)\right) \nonumber\\
% &-& \log_2\!\left(1 - p(d,r) - p(l)\right)\\
% =& 0.\nonumber
%\label{eq:entropyMAX2}
%\end{eqnarray}
we get the solution of $p(d,r)$
\begin{eqnarray}\label{eq:maxEntpq1}
   p(d,r) &=& [1-p(u)][1-p(l)].
   \end{eqnarray}
In the same manner, we get the other three solutions for maximum entropy value as follow
\begin{eqnarray}\label{eq:maxEntpq2}
   p(u,l) &=&  p(u)p(l)    \nonumber\\
   p(u,r) &=&  p(u)[1-p(l)] \\
   p(d,l) &=&  [1-p(u)]p(l). \nonumber
\end{eqnarray}
In short, given a probability distribution of two players' strategies, the prediction of maximum entropy principle is that the joint probability distribution should conform closely to the Eq.~(\ref{eq:maxEntpq1}) and Eq.~(\ref{eq:maxEntpq2}). In another word, the probability distribution of variable $X$ indicating the individual stage strategies must be the function of $p(u)$ and $p(l)$ as it is shown in Eq.~(\ref{eq:maxEntpq1}) and Eq.~(\ref{eq:maxEntpq2}).

Take Eq.~(\ref{eq:maxEntpq1}) and Eq.~(\ref{eq:maxEntpq2}) into Eq.~(\ref{eq:entropydefine}), one can get the maximum entropy in the context of given  $p(u)$ and $p(l)$. It allows us to test the principle of maximum entropy by using economics experimental data.

\section{4 Data set and experiment design}

Conventional experimental methods are well suited to testing whether theories are false. But there are two remaining questions. First, is the sample sufficient? i.e., could  one or two game experiments confirm or reject a theory? Second, is the game carefully selected taking into the experimenter's effect? To avoid these queries, Erev et.al.~\cite{RothErev2007} designed and conducted 40 two-player two-action constant sum games by using random sample to compare different learning models. Among these experimental games, there were 10 games that each had 9 independent repetitive sessions and the others only had one session. Therefore, we selected these 10 games as samples and reorganized them as in Tab.~\ref{tab:RothPayoffEntropy} according to the Table \Rmnum{1} in their paper~\cite{RothErev2007}.

\begin{table}[htbp2]
\begin{threeparttable}
\small
\caption{\label{tab:RothPayoffEntropy} Payoff Matrix of 10 game samples}
\centering
\begin{tabular}{cccccccc}
  \hline
   \hline
  & game\tnote{$\dag$}~~& ~~~~~~$UL$\tnote{$\ddag$}~~~~~~&~~~~~~$UR$~&~~~~~~$DL$~&~~~~~~$DR$~~& ~~~~\\
  \hline
  & 1   & $77$&~~~~~~~~~~ $35$~~~~~~~ & $~8$&~ $48$  \\
  & 2   & $73$&~~~~~~~~~~$74$~~~~~~~ & $87$&~ $20$  \\
  & 3   & $63$&~~~~~~~~~~$~8$~~~~~~~ & $~1$&~ $17$  \\
  & 4   & $55$&~~~~~~~~~~$75$~~~~~~~ & $73$&~ $60$  \\
  & 5   & $~5$&~~~~~~~~~~$64$~~~~~~~ & $93$&~ $40$  \\
  & 6   & $46$&~~~~~~~~~~$54$~~~~~~~ & $61$&~ $23$  \\
  & 7   & $89$&~~~~~~~~~~$53$~~~~~~~ & $82$&~ $92$  \\
  & 8   & $88$&~~~~~~~~~~ $38$~~~~~~~ & $40$&~ $55$  \\
  & 9   & $40$&~~~~~~~~~~ $76$~~~~~~~ & $91$&~ $23$  \\
  & 10  & $69$&~~~~~~~~~~ $~5$~~~~~~~ & $13$&~ $33$  \\
 \hline
  \hline
\end{tabular}
\begin{tablenotes}
  \item[$\dag$] The game 31 to 40 in~\cite{RothErev2007} is relabeled as 1 to 10 in this paper.
  \item[$\ddag$] UL, UR, DL and DR, is labeled as AA, AB, BA and BB respectively.
  \end{tablenotes}
\end{threeparttable}
\end{table}
Each player was asked to select between $A$ and $B$. In accordance with this paper, we call $A$ and $B$ as $U$ and $D$ for player 1, $L$ and $R$ for player 2 respectively. The payoff entry $(i, j)$ presents the probability
$(\times100)$ that player 1 wins when she chose $i$ and her opponent chose $j$ . The payoff for each win was 4 cents which means constant sum game. All 10 games were played by fixed pairs for 500 trials. The right hand columns show the proportion of $U$ choices and $L$ choices in 500 trial by player 1 and player 2 respectively.

For example, if in a given period both players choose action "A" then player 1 will win $v$ with the specified probability $p1$ listed in column $UL$, and player 2 will win $v$ with probability $(1-p1)$. A player who does not win $v$ earns zero in that period. A player's payoff from the game was the sum of his payoffs over the 500 rounds of play (plus a fixed showup fee). Each player played only one game, against a fixed opponent. All transactions were conducted anonymously via networked computers in U.S.. In each random sample described below, the probabilities $p1$ through $p4$ were independently chosen from the uniform distribution on the values $[0.00, 0.01, . . . 0.99, 1.00]$. Games generated in this way were included in the sample if they had a unique mixed strategy equilibrium.

\begin{figure}
\centering
\includegraphics[angle=0,width=3.5cm]{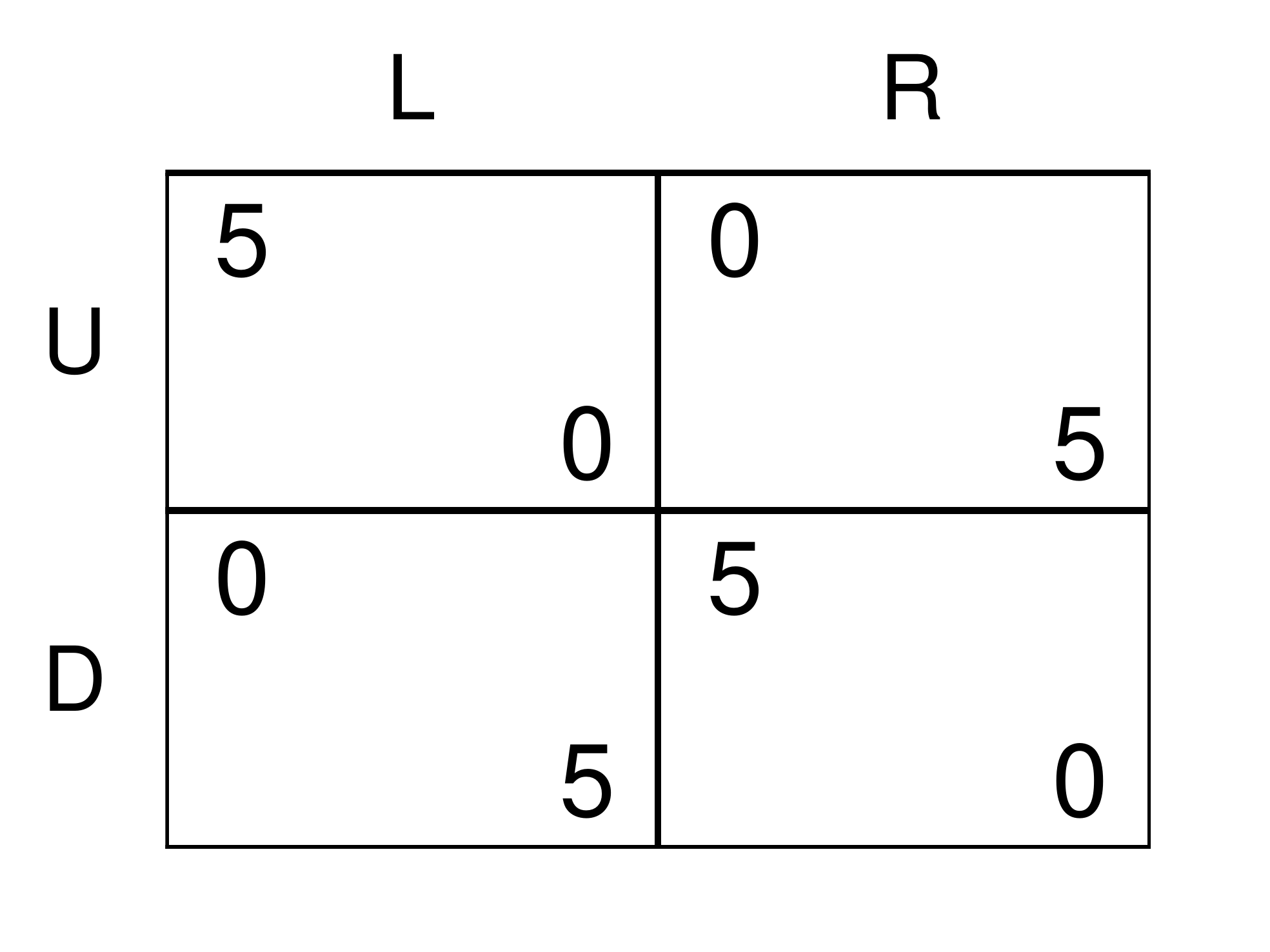}
\caption{Payoff Matrix of matching pennies game.
\label{fig:payoffmatrix4}}
\end{figure}

For testing the principle of maximum entropy in an extreme situation , we plus an extra matching pennies game with payoff matrix in Fig.~\ref{fig:payoffmatrix4} using the form as in Fig.~\ref{fig:payoffmatrix}, for which the theoretical predication entropy should reach the maximum value 2.

%\begin{table}[htbp2]
%\small
%\caption{\label{tab:Payoff Matrix of mp} Payoff Matrix of game11}
%\centering
%\begin{tabular}{cccccccc}
%
%  &           &&        &player&2 \\
%
%\hline
%  &           &&       & ~~$L$~~     &~~ $R$~~ \\
%  & player 1  &&   $U$ & $~5,~0$ & $~0,~5$ \\
%  &           &&   $D$ & $~0,~5$ & $~5,~0$ \\
%\hline
%\end{tabular}
%\end{table}

We conducted the experiment called game 11 similar to the 10 games in Dec. 14, 2010 in Zhejiang University in China with 12 independent pairs for 300 trials. Total 24 subjects participated in this game through recruit. The payoff is 0.35 yuan RMB per 5 points, added fixed showing up fee 5 yuan RMB. The $p(u)$ and $p(l)$ comes out to be $0.512$ and $0.514$, respectively, closing to the the Mixed Strategy Nash Equilibrium $(0.5,0.5)$.

\section{5 result}

\textbf{Results on game level}

There are 9000 decision-making ${(2\times500\times9)}$ in each of the constant sum game from game 1 to game 10 each of which consists of 9 pairs of subjects, each pair play 500 rounds and 7200 decision-making ${(2\times300\times12)}$ in game 11 which consists of 12 pairs of subjects, each pair play 300 rounds. These turn out 4500 observations of strategy pair as variable $X$ in each of game 1 to game 10 and 3600 observations of strategy pair in game 11.

Looking at the two types of players separately, those in the role of player 1 who chose $U$ and those in the role of player 2 who chose $L$ in each game are described by $p(u)$ and $p(l)$ in the second and third column of Tab.~\ref{tab:Entropyin11constantsumgames}. The predictive maximum entropy for each of the 11 games is calculated according to the Eq.~(\ref{eq:maxEntpq2}) and Eq.~(\ref{eq:entropydefine}) under the probability distribution $p(u)$ and $p(l)$ in each game. The results named as maximum entropy(MaxEnt) are shown in the fourth column in Tab.~\ref{tab:Entropyin11constantsumgames}. Similarly, we compute the experimental entropy named observation of entropy(ObEnt) according to the joint probability from raw data by Eq.~(\ref{eq:entropydefine}). In the text, we call the ObEnt as the empirical entropy and lay out the results explicit in the last column in Tab.~\ref{tab:Entropyin11constantsumgames}.

\begin{table}[htbp2]
\begin{threeparttable}
\small
\caption{\label{tab:Entropyin11constantsumgames} Entropy in 11 constant sum games }
\centering
\begin{tabular}{cccccccc}
  \hline
   \hline
  &~game&~~$p(u)$~&~~$p(l)$~~&~~MaxEnt\tnote{$\dag$}~ &~~ObEnt\tnote{$\ddag$}~ &~~obs\tnote{$\S$}~~~ \\
  \hline
  &~ 1  &  $~0.591$ & $~0.318$ & 1.877& 1.862   &4500\\
  &~ 2  &  $~0.840$ & $~0.360$ & 1.576& 1.577	&4500\\
  &~ 3  &  $~0.583$ & $~0.222$ & 1.730& 1.738	&4500\\
  &~ 4  &  $~0.274$ & $~0.502$ & 1.847& 1.847	&4500\\
  &~ 5  &  $~0.378$ & $~0.320$ & 1.861& 1.862	&4500\\
  &~ 6  &  $~0.638$ & $~0.410$ & 1.921& 1.920	&4500\\
  &~ 7  &  $~0.295$ & $~0.522$ & 1.870& 1.849   &4500\\
  &~ 8  &  $~0.400$ & $~0.226$ & 1.741& 1.740   &4500\\
  &~ 9  &  $~0.562$ & $~0.449$ & 1.981& 1.981   &4500\\
  &~ 10 &  $~0.320$ & $~0.202$ & 1.624& 1.609   &4500\\
  &~ 11 &  $~0.512$ & $~0.514$ & 1.999& 1.999	&3600\\
 \hline
 \hline
  \end{tabular}
  \begin{tablenotes}
  \item[$\dag$] The abbreviation of theatrical predicted maximum entropy under the given probability distribution.
  \item[$\ddag$] The abbreviation of observed entropy in experiments.
  \item[$\S$] The total observed frequencies of any probability in each round of experiments.
  \end{tablenotes}
   \end{threeparttable}
  \end{table}

The values of observed entropy are very close to the predictions of the maximum entropy. The mean difference (observed entropy minus maximum entropy)  is $-0.004$, standard deviation is $0.009$, minimum difference is $-0.021$ and the max is 0.008. Ratio estimation of ObEnt/MaxEnt is linearized ratio equal to 0.998, the standard error is $0.001$, $95\% $ confidence interval is $[0.995, 1.001]$.

\begin{figure}
\centering
\includegraphics[angle=0,width=6cm]{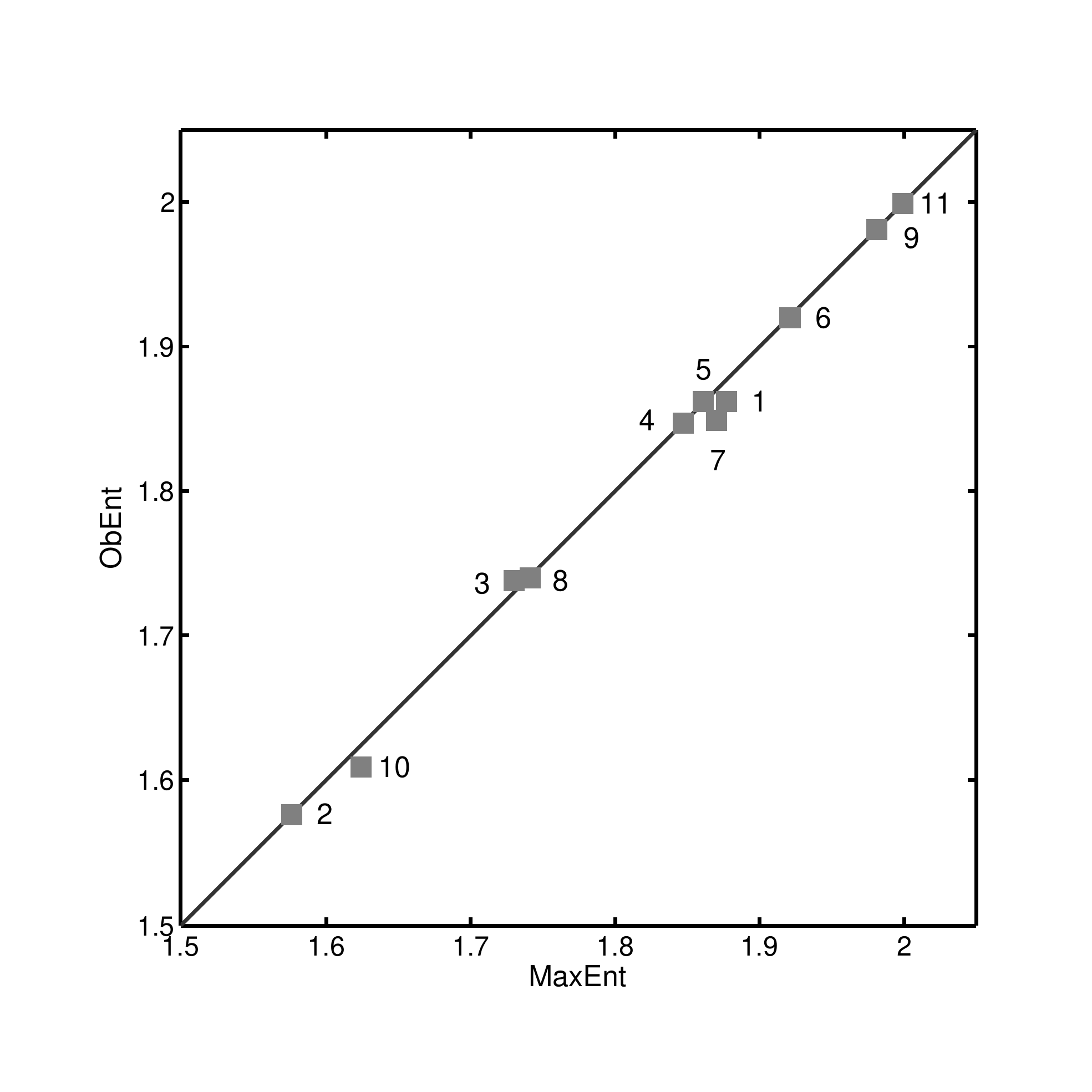}
\caption{Obtained Entropy in the eleven 2 $\times$ 2 game with mixed strategy Nash Equilibrium. Experimental data comes from Ref.~\cite{RothErev2007} and our experiment. Horizon axis is the theoretical expected value of the maximum entropy hypothesis and the vertical axis is the empirical value.}
\label{fig:EntropySeltenRothTheoMeasuredTestNormalDistribution}
\end{figure}

Fig.~\ref{fig:EntropySeltenRothTheoMeasuredTestNormalDistribution} plots the results of comparing theoretical value of maximum entropy and empirical value of entropy in the 11 experiments. The theoretical value is expressed in horizon axis and the empirical value is in vertical axis. Each square with a numerical game label denotes the mean use of $U$ strategy and $L$ strategy in that game. A diagonal line which implies the empirical value equivalent to theoretical value added for revealing how empirical entropy approximate to the maximum entropy. The result of linear regression is in Tab.~\ref{tab:resultoflinearregression}.

\begin{table}[htbp2]
\begin{threeparttable}
\small
\caption{\label{tab:resultoflinearregression} result of linear regression }
\centering
\begin{tabular}{cccccccc}
  \hline
   \hline
  &obent	&Coef.	&Std.Err.	&~~t~~  &$P>|t|$	&[95\%Conf.Interval]	\\
  \hline
  &maxent	&~1.000	&0.022	    &46.36	&~~~~0	    &~0.952~~~1.049\\
  & cons	&-0.005	&0.039	    &-0.13	&0.901	    &-0.094~~~0.084\\
 \hline
 \hline
  \end{tabular}
     \end{threeparttable}
  \end{table}

The results on game level show, the constant sum game fits the Principle of Maximum Entropy very well.

\textbf{Results on group level}

Now, we will report the results at group level.

\begin{table}[htbp2]
\begin{threeparttable}
\small
\caption{\label{tab:Entropyongrouplevel} Entropy on group level }
\centering
\begin{tabular}{cccccccc}
  \hline
   \hline
  &~~~~~&~~~~~game1~~~~&~~~~~game2~~~~&~~~~game3~~~~~ \\
  &group&~MaxEnt~ObEnt~&~MaxEnt~ObEnt~&~MaxEnt~ObEnt~ \\
   \hline
  & 1   &~$1.407$~~~$1.407$&~$1.773$~~~$1.752$&~$1.817$~~~$1.808$ \\
  & 2   &~$1.947$~~~$1.942$&~$0.436$~~~$0.436$&~$1.311$~~~$1.311$  \\
  & 3   &~$1.965$~~~$1.956$&~$1.974$~~~$1.974$&~$0.985$~~~$0.985$ \\
  & 4   &~$1.830$~~~$1.829$&~$1.474$~~~$1.469$&~$1.653$~~~$1.652$  \\
  & 5   &~$1.965$~~~$1.519$&~$1.137$~~~$1.136$&~$1.452$~~~$1.416$  \\
  & 6   &~$1.763$~~~$1.763$&~$1.860$~~~$1.820$&~$1.671$~~~$1.670$ \\
  & 7   &~$1.000$~~~$0.781$&~$1.036$~~~$1.036$&~$1.677$~~~$1.512$  \\
  & 8   &~$1.713$~~~$1.711$&~$1.500$~~~$1.495$&~$1.363$~~~$1.362$  \\
  & 9   &~$1.320$~~~$1.314$&~$1.441$~~~$1.441$&~$1.403$~~~$1.388$  \\
  \hline
  \hline
  &~~~~~&~~~~~game4~~~~&~~~~~game5~~~~&~~~~game6~~~~~ \\
  &group&~MaxEnt~ObEnt~&~MaxEnt~ObEnt~&~MaxEnt~ObEnt~ \\
  \hline
  & 1   &~$1.273$~~~$1.273$&~$1.984$~~~$1.963$&~$1.776$~~~$1.775$ \\
  & 2   &~$1.886$~~~$1.842$&~$1.707$~~~$1.695$&~$1.992$~~~$1.992$  \\
  & 3   &~$1.765$~~~$1.727$&~$1.393$~~~$1.343$&~$1.951$~~~$1.938$ \\
  & 4   &~$0.938$~~~$0.931$&~$1.922$~~~$1.910$&~$1.941$~~~$1.928$  \\
  & 5   &~$1.992$~~~$1.970$&~$1.888$~~~$1.888$&~$1.872$~~~$1.868$  \\
  & 6   &~$1.474$~~~$1.473$&~$1.760$~~~$1.754$&~$1.514$~~~$1.438$ \\
  & 7   &~$1.860$~~~$1.835$&~$1.932$~~~$1.928$&~$1.960$~~~$1.959$  \\
  & 8   &~$1.781$~~~$1.738$&~$1.704$~~~$1.702$&~$1.799$~~~$1.782$  \\
  & 9   &~$1.993$~~~$1.937$&~$1.640$~~~$1.638$&~$1.864$~~~$1.864$  \\
  \hline
  \hline
  &~~~~~&~~~~~game7~~~~&~~~~~game8~~~~&~~~~game9~~~~~ \\
  &group&~MaxEnt~ObEnt~&~MaxEnt~ObEnt~&~MaxEnt~ObEnt~ \\
  \hline
  & 1   &~$1.000$~~~$1.000$&~$1.027$~~~$1.026$&~$1.955$~~~$1.950$ \\
  & 2   &~$1.857$~~~$1.843$&~$1.579$~~~$1.539$&~$1.961$~~~$1.959$  \\
  & 3   &~$1.701$~~~$1.696$&~$1.661$~~~$1.658$&~$1.935$~~~$1.890$ \\
  & 4   &~$1.876$~~~$1.875$&~$1.832$~~~$1.790$&~$1.970$~~~$1.915$  \\
  & 5   &~$1.862$~~~$1.758$&~$1.853$~~~$1.806$&~$1.652$~~~$1.651$  \\
  & 6   &~$1.346$~~~$1.168$&~$1.677$~~~$1.673$&~$1.997$~~~$1.985$ \\
  & 7   &~$1.513$~~~$1.500$&~$0.975$~~~$0.975$&~$1.983$~~~$1.980$  \\
  & 8   &~$1.844$~~~$1.842$&~$1.706$~~~$1.703$&~$1.639$~~~$1.536$  \\
  & 9   &~$1.992$~~~$1.991$&~$1.812$~~~$1.810$&~$1.846$~~~$1.818$  \\
  \hline
   \hline
  &~~~~~&~~~~~game10~~~~&~~~~~game11~~~~&~~~~~~~~~ \\
  &group&~MaxEnt~ObEnt~&~MaxEnt~ObEnt~&~~ \\
  \hline
  & 1   &~$1.620$~~~$1.595$&~$1.996$~~~$1.989$& \\
  & 2   &~$1.087$~~~$1.012$&~$1.997$~~~$1.993$& \\
  & 3   &~$1.822$~~~$1.812$&~$1.996$~~~$1.996$& \\
  & 4   &~$1.579$~~~$1.578$&~$1.989$~~~$1.983$& \\
  & 5   &~$0.775$~~~$0.775$&~$1.999$~~~$1.996$& \\
  & 6   &~$1.945$~~~$1.942$&~$2.000$~~~$1.997$& \\
  & 7   &~$1.910$~~~$1.910$&~$1.981$~~~$1.964$& \\
  & 8   &~$1.423$~~~$1.420$&~$1.999$~~~$1.998$& \\
  & 9   &~$0.600$~~~$0.600$&~$1.995$~~~$1.994$& \\
  & 10  &~$     $~~~$     $&~$1.999$~~~$1.998$& \\
  & 11  &~$     $~~~$     $&~$1.995$~~~$1.994$& \\
  & 12  &~$     $~~~$     $&~$1.998$~~~$1.993$& \\
  \hline
 \hline
  \end{tabular}
     \end{threeparttable}
  \end{table}

There are 9 independent groups with two fixed players in each game from game 1 to game 10, and 12 independent groups with two fixed players in game 11. Each group can be looked as an independent social interaction system. We can test the principle on group level in each game too. We calculate the theoretical maximum entropy and empirical entropy in each group, the results explicitly lay in Tab.~\ref{tab:Entropyongrouplevel}. Each 9 groups from game 1 to game 10 and 12 groups from game 11 have paired theoretical maximum entropy and empirical entropy.

Interesting and natural, no empirical value exceeds the theoretical value. The statistical ratio $ObEnt/MaxEnt$ is exhibited in Tab.~\ref{tab:ratio} which is the mean of the total pairs within a game. The ratio for game 1 to game 11 is $95.4\%$, $99.4\%$, $98.3\%$, $98.4\%$, $99.3\%$, $99.3\%$,  $97.9\%$, $99.0\%$, $98.5\%$, $99.1\%$, $99.8\%$, respectively. All empirical entropy values are near to the theoretical maximum entropy values, no one is inferior to $95\%$.

\begin{table}[htbp2]
\begin{threeparttable}
\small
\caption{\label{tab:ratio} Ratio of $ObEnt/MaxEnt$ }
\centering
\begin{tabular}{cccccccc}
  \hline
   \hline
  game&~$~obs$~&$Ratio$~&~$Std. Err.~$ &~$[95\% Conf. Interval]~$\\
   \hline
   1   &$~9$~&~$0.954$&~$0.031$&~~$0.881$~~~~~~1.026~~ \\
   2   &$~9$~&~$0.994$&~$0.003$&~~$0.987$~~~~~~1.001~~ \\
   3   &$~9$~&~$0.983$&~$0.012$&~~$0.956$~~~~~~1.010~~ \\
   4   &$~9$~&~$0.984$&~$0.003$&~~$0.976$~~~~~~0.992~~ \\
   5   &$~9$~&~$0.993$&~$0.003$&~~$0.986$~~~~~~1.000~~ \\
   6   &$~9$~&~$0.993$&~$0.004$&~~$0.982$~~~~~~1.002~~ \\
   7   &$~9$~&~$0.979$&~$0.013$&~~$0.949$~~~~~~1.009~~ \\
   8   &$~9$~&~$0.990$&~$0.004$&~~$0.981$~~~~~~1.000~~ \\
   9   &$~9$~&~$0.985$&~$0.006$&~~$0.971$~~~~~~1.000~~ \\
   10  &$~9$~&~$0.991$&~$0.006$&~~$0.977$~~~~~~1.004~~ \\
   11  &$~12$~&~$0.998$&~$0.001$&~~$0.996$~~~~~~0.999~~ \\
  \hline
 \hline
  \end{tabular}
   \end{threeparttable}
  \end{table}

\section{6 discuss and conclusion}

By employing experimental economics data, we test the \emph{Principle of Maximum Entropy} hypothesis. The results show, the entropy in constant sum game system fits the Principle of Maximum Entropy very well. Thereby, we firstly test the maximum entropy principle in social interaction system with the simplest and competitive situation in laboratory economics experiments.

However, notice that the data used here all come from the constant sum games with theoretical Mixed Strategy Nash Equilibrium. Does this principle apply to other games, i.e., social systems with other situations?

First, as we know,  because of existing Mixed Strategy Nash Equilibrium in these games, there should be some uncertainty within it. Therefore, the degree of fitness with principle of maximum entropy may come out to be a signal for estimating the type of a game, i.e., the uncertainty degree within a game. Second, in physics, the maximum entropy principle is established on the isolation of the system. That is to say, there are no exchange of energy or matter with environment, if there is some energy exchange with environment, the condition of maximum entropy principle has to be destroyed. Perhaps, there exist some relationships between economics and physics. As we mentioned these games are all constant sum game, which means each payoff sum of four outcome states is equivalent.

For instance, in their control sessions which provide entropy result in their paper, Bednary et.al.~\cite{Yan2011} implement four type games named $PD$, $SA$, $WA$ and $SI$ respectively. $PD$ means prisoner's dilemma game, $SA$ and $WA$ means strange and weak alternative game, and the SI means the selfish interesting game.

\begin{table}[htbp2]
\begin{threeparttable}
\small
\caption{\label{tab:EntropyinBednarygame} Entropy in Ref.~\cite{Yan2011,myerson1983mechanism} games}
\centering
\begin{tabular}{cccccccccc}
   \hline
   \hline
     &     & Payoff Matrix\tnote{$\dag$}&&ProbDistr\tnote{$\ddag$}&ObEnt\tnote{$\S$}&MaxEnt\tnote{$\natural$}\\
    \hline
     &     & $~C$~~~~~~$S$    &&$C~~~~$~~$S$    &                  \\
     \hline
     &~~$C$& $~7,~7$ ~$~2,10$&&~$55.68$~~$11.67$&                  \\
  PD\tnote{$\pounds$} &~~$S$& $10,~2$ ~$~4,~4$&&~$14.82$~~$17.82$&$1.68$&1.79  \\
  \hline
     &~~$C$& $~7,~7$ ~$~4,14$&&~$~5.02$~~$39.81$&                  \\
  SA\tnote{$\pounds$} &~~$S$& $14,~4$ ~$~5,~5$&&~$40.37$~~$14.81$&$1.68$&1.99  \\
  \hline
     &~~$C$& $~7,~7$ ~$~4,11$&&~$33.18$~~$21.57$&                  \\
  WA\tnote{$\pounds$} &~~$S$& $11,~4$ ~$~5,~5$&&~$22.74$~~$22.51$&$1.98$&1.98  \\
     \hline
     &~~$C$& $~7,~7$ ~$~2,~9$&&~$~0.00$~~$~0.14$&                  \\
  SI\tnote{$\pounds$} &~~$S$& $~9,~2$ ~$10,10$&&~$~0.00$~~$99.86$&$0.02$&0.02  \\
     \hline
     \hline
     &~~$1a$& $~5,~~3$ ~$~9,-2$&&~$~20.00$~~$~30.00$&                  \\
  M1\tnote{$\pounds$} &~~$1b$& $~9,-2$ ~$~5,~~3$&&~$~30.00$~~$~20.00$&$1.97$&2.00  \\
     \hline
     \hline
    \end{tabular}
     \begin{tablenotes}
  \item[$\dag$] The payoff matrix for $PD$, $SA$, $WA$ and $SI$ respectively, where $C$ means corporate and $S$ means selfish or defect. The payoff entry $(i, j)$ presents the payoff when Player 1 chose $i$ and her opponent chose $j$ .
  \item[$\ddag$] The abbreviation of probability distribution, the numerical  presents the probability $(\times100)$ that Player 1 chose $i$ and her opponent chose $j$ .
  \item[$\S$] The abbreviation of observed entropy in experiment.
   \item[$\natural$] The abbreviation of theatrical predicted maximum entropy under the given probability distribution.
   \item[$\pounds$] Game $PD$, $SA$, $WA$ and $SI$ are from \cite{Yan2011} and game $M1$ is from \cite{myerson1983mechanism}.
  \end{tablenotes}
\end{threeparttable}
\end{table}

The results are reorganized and exhibited in Tab.~\ref{tab:EntropyinBednarygame}. Followed the protocol of infinitely repeated games in the laboratory, the game they implemented is  an infinitely repeated game, with a discount factor of 1 for the first 200 rounds, and 0.9 thereafter, one 12-player session for each of the single games. Participants are randomly matched into pairs at the beginning of each session, and play the same match for the entire experiment. We calculate the theoretical maximum entropy for each game according to the $p(u)$ and $p(l)$ in their games. Distinguish with the previous results, the observed entropy here seems to be something different from the entropy in constant sum game. For example, the entropy in $PD$ game and $SA$ game is equal to $1.68$ while the prediction of maximum entropy is 1.788 and 1.986 respectively, though the entropy in $WA$ game and $SI$ game is near to its' maximum entropy prediction. Looking at the payoff matrix carefully, all the games here are not the constant sum game.

Another example for application of MaxEnt comes from incomplete information games suggested by Myerson~\cite{myerson1983mechanism} for mechanism design. The last row in Table~\ref{tab:EntropyinBednarygame} in this paper is the payoff matrix of the \textit{example 1} in~\cite{myerson1983mechanism}. Theoretically, the mean observation should be (0.5,~0.5), so the MaxEnt should be 2. However, due to the asymmetry of information, the strategy of the informed principal player (the row player) should depend on the strategy of the uninformed column player. Because row player has the opponent's strategy information, so he/she can control the result of the game, at the same time, he/she should be constrained by an incentive-compatible mechanism and must give nonnegative expected payoff to the uninformed column player (for more details of the constraint condition, see~\cite{myerson1983mechanism}). Supposing that the uninformed column player indeed chooses the two strategies equally, the row player should have 60\% opportunity to win; as a consequence, the distribution of the observation should not spread equally in the 4 states, and the observed entropy should be lower than 2, see Table~\ref{tab:EntropyinBednarygame}. In this case, we show how to evaluate the symmetry of private information in a bargain with the MaxEnt.

In economics, the Nash Equilibrium means a stable state, in which nobody is willing to change the state. In physics, a system will reach its' stable state if the entropy increases to its' maximum value. The relationship  between these two stable state notions seems to be interesting and deserving further studying. Of course, this needs more economics experiments and cooperative study.

\textbf{Acknowledgment:} Thanks to Timothy N. Cason, Yan Chen and Alvin Roth for the data support.

\bibliography{mainMar0721forHEn}

%\bibliography{mainMar0517}

\end{document}